\documentclass[twocolumn]{IEEEtran}
\usepackage{cite}
\usepackage{amsmath,epsfig,amssymb,verbatim,amsopn,cite,multirow}
\usepackage{amsthm}
\usepackage{balance}
\usepackage{multirow}
\usepackage[usenames,dvipsnames]{color}
\usepackage[all]{xy}  %%% used to make block diagram
\usepackage{url}
\usepackage{amsfonts}
\usepackage{amssymb}
\usepackage{epsfig}
\usepackage{epstopdf}
\usepackage{bm}
\usepackage{balance}
\usepackage{graphicx}
\usepackage{subcaption}
\usepackage{footnote}
\usepackage{cancel}
\usepackage{algorithm,algorithmic}
\usepackage{balance}
\usepackage[margin=14.5mm,top=18.1mm,bottom=25.7mm]{geometry}%margins; tweak for printer deficiencies

\usepackage[nodisplayskipstretch]{setspace}
\newcommand{\qa}{{\bf a}}

\newcommand{\qc}{{\bf c}}

\newcommand{\qe}{{\bf e}}

\newcommand{\qg}{{\bf g}}
\newcommand{\qh}{{\bf h}}

\newcommand{\qn}{{\bf n}}

\newcommand{\qu}{{\bf u}}

\newcommand{\qw}{{\bf w}}
\newcommand{\qx}{{\bf x}}
\newcommand{\qy}{{\bf y}}
\newcommand{\qz}{{\bf z}}

\newcommand{\qA}{{\bf A}}
\newcommand{\qB}{{\bf B}}

\newcommand{\qD}{{\bf D}}
\newcommand{\qE}{{\bf E}}

\newcommand{\qH}{{\bf H}}
\newcommand{\qI}{{\bf I}}

\newcommand{\qN}{{\bf N}}

\newcommand{\qP}{{\bf P}}

\newcommand{\qY}{{\bf Y}}

\newcommand{\Ex}{\mathbb{E}}
\newcommand{\SSE}{\mathrm{SSE}}
\newcommand{\SE}{\mathrm{SE}}

\newcommand{\UEk}{\mathrm{UE}_k}

\newcommand{\UEo}{\mathrm{UE}_1}
\newcommand{\SINR}{\mathrm{SINR}}

\begin{document}
\title{{Toward Security-Resilient Cell-Free Massive MIMO: A Multi-Stage Framework}
\thanks{}}

\author{
    \IEEEauthorblockN{Junbin Yu, Tianyu Lu, Mohammadali Mohammadi, and Michail Matthaiou}
    \IEEEauthorblockA{
        \\ Centre for Wireless Innovation (CWI), Queen’s University Belfast, Belfast, BT7 1NN, U.K. \\
        E-mail: \{jyu17, t.lu, m.mohammadi, m.matthaiou\}@qub.ac.uk}
        \thanks{This work was supported by the European Research Council (ERC) under the European Union’s Horizon 2020 research and innovation programme (grant agreement No. 101001331), UK Engineering and Physical Sciences
Research Council (EPSRC) grant EP/X04047X/2 for TITAN Telecoms Hub, and by a research grant from the Department for the Economy Northern Ireland under the US-Ireland R\&D Partnership Programme.}}

\maketitle
\begin{abstract}
This paper develops a robust security-resilient transmission framework for cell-free massive multiple-input multiple-output (CF-mMIMO) systems under active pilot spoofing attacks. As a baseline, system performance is characterized under attack-free conditions to establish the target user’s pre-attack service level. Upon attack detection, the system enters an absorption phase, during which, power allocation is adaptively adjusted across a limited subset of access points (APs) using contaminated channel state information (CSI). This phase quickly compensates for performance degradation while maintaining low operational overhead. Once the secrecy spectral efficiency (SSE) recovers to a prescribed loss level, the resulting power allocation initializes the restoration phase. Here, the transmit powers of all APs are jointly optimized, and a protective partial zero-forcing (PPZF) strategy further improves secrecy. In parallel, artificial noise (AN) is incorporated under a worst-case eavesdropping scenario accounting for large-scale fading uncertainty. The resulting stage-dependent non-convex problems are formulated within a unified framework and solved using successive convex approximation (SCA). Numerical results demonstrate that the proposed scheme achieves an effective time–quality tradeoff while maintaining the highest recovered secrecy; in a representative setup, it achieves gains of up to $3.6\%$, $15.7\%$, and $56\%$ over the respective baseline schemes, with similar improvements under eavesdropper’s channel uncertainty.
\end{abstract}

%----------------------
%vspace{-1em}

%----------------------
\vspace{-1.em}
\section{Introduction}
%--------------------------
\IEEEPARstart{C}{F}-mMIMO is a promising wireless architecture thanks to its enhanced macro-diversity gain, user-centric service, and improved coverage uniformity \cite{mohammadi2024next}. By allowing distributed APs to cooperatively serve users, CF-mMIMO enhances spectral efficiency (SE) and reliability \cite{hien}. However, it faces various security threats, including passive eavesdropping, jamming, and active attacks that exploit the channel acquisition process \cite{matthaiou2021road}.

Among these threats, active pilot spoofing is particularly harmful: during uplink training, an eavesdropper (Eve) transmits the target user’s pilot, contaminating its channel estimate. Consequently, downlink transmission is partly steered toward Eve, increasing information leakage and degrading secrecy~\cite{Kapetanovic:CM:2015}. Existing CF-mMIMO physical layer security (PLS) designs enhance secrecy through pilot assignment, power allocation, AP clustering, or AN~\cite{Gayan:GC:2018,Atiya:TWC:2024,Chen:IoT:2024,Banik:TWC:2026}. \textit{However, they largely overlook processing latency and leave a critical question unanswered: ``How can the system respond immediately to an attack and restore secrecy under persistently contaminated CSI?''}

% Among these threats, an active pilot spoofing attack is particularly detrimental. In this attack, an eavesdropper (Eve) transmits the same pilot sequence as a target user during uplink training, thereby contaminating the channel estimate. As a result, the downlink transmission may be partially steered toward Eve, leading to increased information leakage and degraded secrecy performance for the legitimate user ~\cite{Kapetanovic:CM:2015}. Existing physical layer security (PLS) designs for CF-mMIMO mainly focus on enhancing secrecy under fixed operating conditions, for example, through pilot assignment, power allocation, AP clustering, or artificial noise (AN)-aided transmission~\cite{Gayan:GC:2018,Atiya:TWC:2024,Chen:IoT:2024,Banik:TWC:2026}. \textit{However, these works often overlook processing latency in practical systems—specifically, they have not answered the question: ``How to react immediately after an attack and how to restore secrecy performance under persistently contaminated CSI?''}

This limitation motivates a resilience view of secure transmission~\cite{ICC_resilient,Jamming_resilience}. Resilience can be interpreted as an adaptive response to disruptions, where recovery actions are adjusted to absorb performance degradation and restore a feasible system trajectory~\cite{resilience-CDC}. Recent studies have investigated resilience for post-disruption resource management and service recovery~\cite{Reifert1}, and have further incorporated absorption–adaptation-based resilience modeling into secure wireless systems with two-timescale recovery designs~\cite{Wu_resilience}. Different from these works, we propose a robust security-resilient transmission framework that embeds resilience into the post-attack physical-layer control of CF-mMIMO under an active pilot spoofing attack. In this framework, the system first absorbs the immediate secrecy degradation, through a low-overhead response, and then restores the secure operating point under contaminated CSI. The main contributions are summarized as follows:
%\vspace{-1em}
\begin{itemize}
    \item We propose a robust security-resilient transmission framework for CF-mMIMO systems under active pilot spoofing, which separates long-term normal benchmarking from short-term post-attack absorption and restoration with stage-dependent action spaces.  
    \item We propose a multi-stage recovery framework under contaminated CSI. In the absorption stage, low-overhead local power reallocation over a small AP subset is optimized, providing rapid secrecy loss mitigation and a feasible warm start for later recovery. In the restoration stage, AN and PPZF reconfiguration (PR) are used to optimize power allocation across the network.
    \item We formulate a worst-case secure transmission model under large-scale fading uncertainty for Eve and solve the resulting non-convex normal-stage, absorption-stage, and restoration-stage designs via a unified SCA-based framework. Numerical results verify the effectiveness of the proposed design in both post-attack mitigation and final secrecy restoration.
\end{itemize}

\textit{Notation:} Boldface lowercase and uppercase letters denote vectors and matrices, respectively. The symbols $(\cdot)^{-1}$, $(\cdot)^\mathrm{T}$ and $(\cdot)^\mathrm{H}$ denote the inverse operator, transpose operator and conjugate transpose operator; the operator $\mathrm{diag}(\qA)$ denotes a vector consisting of the diagonal elements of matrix $\qA$; $\qI_M$ denotes the $M\times M$ identity matrix; $\|\cdot\|$ denotes the Euclidean norm of a vector or the Frobenius norm of a matrix, while $|\cdot|$ represents the absolute value of a scalar; a circular-symmetric complex Gaussian variable having variance $\sigma^2$ is denoted by $\mathcal{CN}(0,\sigma^2)$.  Finally, $\mathbb{E}\{\cdot\}$ denotes the statistical expectation.

%%%%%%%%%%%%%%%%%%%%%%%%%%%%%%%%%%%%%%%%%%%
\vspace{-0.5em}
\section{System Model}
We consider a downlink CF-mMIMO network with $L$ distributed APs and $K$ single-antenna users. Each AP deploys $M$ antennas and is connected to a central processing unit (CPU) through fronthaul links. For simplicity of notation, we define the sets $\mathcal{K}\triangleq\{1,\ldots,K\}$, and $\mathcal{L}\triangleq\{1,\ldots,L\}$ to represent the sets of the users and APs, respectively. Moreover, $\UEk$ denotes the user $k$. A single-antenna active Eve targets $\UEo$ by transmitting the same pilot sequence as that user during uplink training.

%-----------------------------
\vspace{-0.7em}
\subsection{Channels and Pilot Spoofing}
%-----------------------
Let $\qh_{l,k}\in\mathbb{C}^{M\times 1}$ and $\qh_{l,e}\in\mathbb{C}^{M\times 1}$ denote the channel vectors from the $l$-th AP to $\UEk$ and Eve. We consider independent Rayleigh fading channels, which are modeled as $\qh_{l,k}=\sqrt{\beta_{l,k}}\,\qg_{l,k}, \qh_{l,e}=\sqrt{\beta_{l,e}}\,\qg_{l,e}$,
where $\beta_{l,k}$ and $\beta_{l,e}$ denote the large-scale fading coefficients, while $\qg_{l,k},\,\qg_{l,e}\sim\mathcal{CN}(\boldsymbol{0},\qI_M)$. Equivalently, we have $\qh_{l,k}\sim\mathcal{CN}(\boldsymbol{0},\beta_{l,k}\qI_M)$ and $\qh_{l,e}\sim\mathcal{CN}(\boldsymbol{0},\beta_{l,e}\qI_M)$.

During the uplink training phase, all users simultaneously transmit their pilot sequences to the APs. Let $\boldsymbol{\phi}_k\in\mathbb{C}^{\tau_p\times 1}$ denote the pilot sequence assigned to $\UEk$. We assume that $\tau_p \ge K$, thus the pilot sequences satisfy $\boldsymbol{\phi}_k^\mathrm{H}\boldsymbol{\phi}_{k'}=0, k\neq k'$, and $\|\boldsymbol{\phi}_k\|^2=1$.
Meanwhile, Eve attacks the channel estimation of the target user, $\UEo$, by transmitting the identical pilot sequence, i.e., $\boldsymbol{\phi}_e=\boldsymbol{\phi}_1$.

Accordingly, the received pilot matrix at the $l$-th AP is
\begin{equation}
\qY_{p,l}=\sqrt{\tau_p\rho_p}\sum\nolimits_{k\in\mathcal{K}}\qh_{l,k}\boldsymbol{\phi}_k^\mathrm{H}+\sqrt{\tau_p\rho_e}\qh_{l,e}\boldsymbol{\phi}_1^\mathrm{H}+\qN_l,
\label{eq:pilot_rx_matrix}
\end{equation}
where $\tau_p$ is the pilot length, $\rho_p\triangleq P_p/N_0$ and $\rho_e\triangleq P_e/N_0$ denote the pilot transmit power of the legitimate users and Eve, respectively, with noise power $N_0$. Moreover, $\qN_l\in\mathbb{C}^{M\times \tau_p}$ is the additive noise matrix with independent and identically distributed (i.i.d.) $\mathcal{CN}(0,1)$ elements. Now, by projecting $\qY_{p,l}$ onto $\boldsymbol{\phi}_k$, the $l$-th AP obtains
\begin{equation}
\qy_{l,k}=\qY_{p,l}\boldsymbol{\phi}_k=\sqrt{\tau_p\rho_p}\,\qh_{l,k}+\delta_k\sqrt{\tau_p\rho_e}\,\qh_{l,e}+\qn_{l,k},
\label{eq:pilot_projection}
\end{equation}
where $\delta_k=1$ if $k=1$ and $\delta_k=0$ otherwise. Moreover, $\qn_{l,k}\triangleq \qN_l\boldsymbol{\phi}_k\sim\mathcal{CN}(\boldsymbol{0},\qI_M)$.

Using minimum mean square error (MMSE) estimation, the estimate of $\qh_{l,k}$ is given by~\cite{Atiya:TWC:2024}
\begin{equation}
\hat{\qh}_{l,k}
=
\frac{\sqrt{\tau_p\rho_p}\,\beta_{l,k}}
{\tau_p\rho_p\beta_{l,k}+\delta_k\tau_p\rho_e\beta_{l,e}+1}
\qy_{l,k}.
\label{eq:mmse_estimator}
\end{equation}
Accordingly, the estimate satisfies $\hat{\qh}_{l,k}\sim\mathcal{CN}(\boldsymbol{0},\gamma_{l,k}\qI_M)$, with $\gamma_{l,k}
=
\frac{\tau_p\rho_p\beta_{l,k}^2}
{\tau_p\rho_p\beta_{l,k}+\delta_k\tau_p\rho_e\beta_{l,e}+1}$.
% \begin{equation}
% \gamma_{l,k}
% =
% \frac{\tau_p\rho_p\beta_{l,k}^2}
% {\tau_p\rho_p\beta_{l,k}+\delta_k\tau_p\rho_e\beta_{l,e}+1}.
% \label{eq:gamma_general}
% \end{equation}
Moreover, the channel estimation error is defined as $\tilde{\qh}_{l,k}=\qh_{l,k}-\hat{\qh}_{l,k}$, which is independent of $\hat{\qh}_{l,k}$ and follows $\tilde{\qh}_{l,k}
\sim\mathcal{CN}\!\left(\boldsymbol{0},(\beta_{l,k}-\gamma_{l,k})\qI_M\right)$.

Here, the stage index is denoted by $s\in\{n,a,r\}$, representing the normal, absorption, and restoration stages, respectively. For the normal stage ($s=n$), no spoofing is present and
%-----------------------------
\vspace{-0.2em}
\begin{equation}
\gamma_{l,k}^{(n)}
=
\frac{\tau_p\rho_p\beta_{l,k}^2}
{\tau_p\rho_p\beta_{l,k}+1},\quad k\in\mathcal K.
\label{eq:gamma_normal}
\end{equation}
For the post-attack stages $s\in\{a,r\}$, only the attacked user, i.e., $\UEo$, suffers from pilot contamination, thus
%-----------------------------
\vspace{-0.2em}
\begin{equation}
\gamma_{l,1}^{(a)}=\gamma_{l,1}^{(r)}
=
\frac{\tau_p\rho_p\beta_{l,1}^2}
{\tau_p\rho_p\beta_{l,1}+\tau_p\rho_e\beta_{l,e}+1},
\label{eq:gamma_target_attack_restore}
\end{equation}
while $\gamma_{l,k}^{(a)}=\gamma_{l,k}^{(r)}=\gamma_{l,k}^{(n)}$ for $k\ge 2$. Hence, both post-attack stages operate under the same contaminated CSI, and differ only in the admissible recovery actions.

%============================
\vspace{-1em}
\subsection{Two-Timescale Secure Framework}\label{two-time}
%============================
\begin{figure}[t]
\centering
\includegraphics[width=3 in]{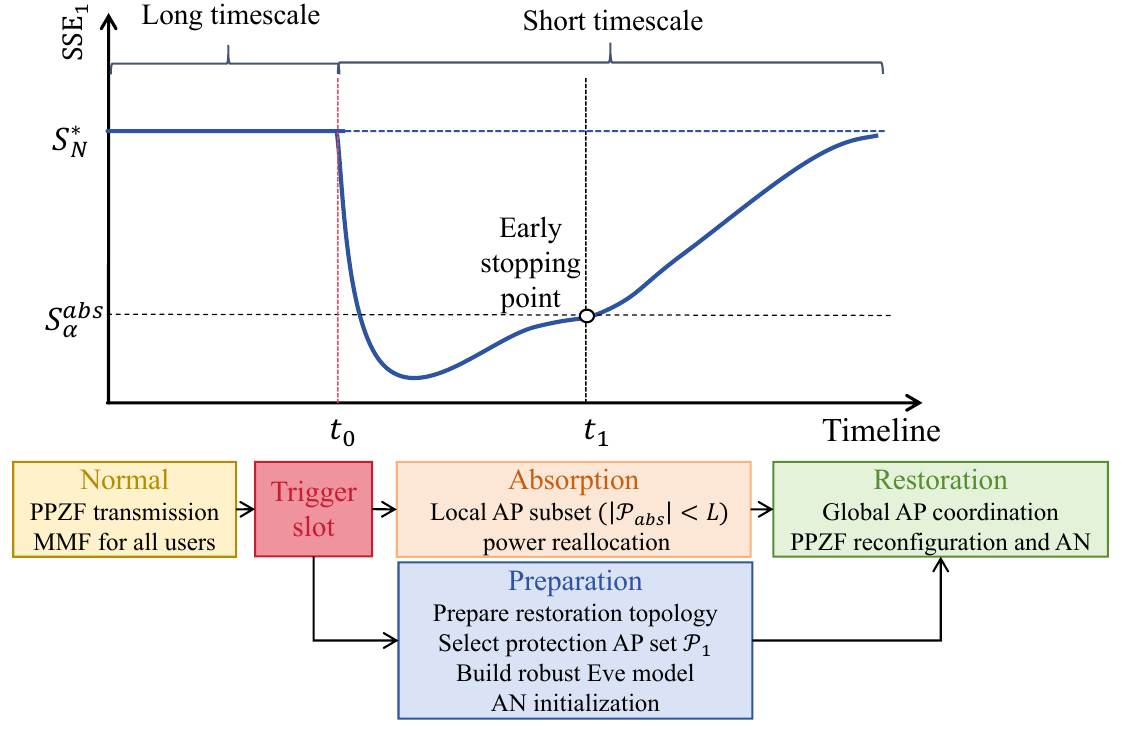}%framework1.png
\vspace{-0.2em}
\caption{Illustration of two-timescale secure resilience framework.}
\label{fig1}
\end{figure}

As shown in Fig.~\ref{fig1}, we consider a secure CF-mMIMO system operated over two timescales. On the long-term scale, the CPU determines a normal benchmark operating point in the absence of attacks. The resulting pre-attack service level of the targeted user is denoted by $S_{\mathrm{N}}^{\star} \triangleq \mathrm{SE}_1^{(n)}\big(\rho_{l,k}^{(n)\star}\big)$, where $\mathrm{SE}_1^{(n)}$ denotes the SE of $\UEo$ in the normal stage and is defined in subsection~\ref{sec:SSE}. %On the short-term scale, once a pilot-spoofing alarm is triggered, the system enters a two-stage post-attack control process with different action spaces:
On the short-term scale, a pilot-spoofing alarm is assumed to be provided by an external detection mechanism, with $t_0$ denoting the instant when the alarm is received by the CPU. The framework focuses on post-detection secure recovery through two stages with different action spaces, while attack detection is left for future work:
\begin{align}
   & \mathcal{A}_{\mathrm{abs}} = \big\{\rho_{l,k}^{(a)}\big\},
~l\in\mathcal L_{\mathrm{abs}},~k\in\mathcal K,\\
&\mathcal{A}_{\mathrm{rec}} = \big\{
\rho_{l,k}^{(r)},~
\rho_{l,\mathrm{AN}}^{(r)},~
\mathcal S_l^{(r)}\big\}, l\in\mathcal L,~k\in\mathcal K. \label{A_rec}
\end{align}
Here, $\rho_{l,k}^{(s)}$ denotes the transmit power coefficient from the $l$-th AP to $\UEk$ in stage $s$, $\rho_{l,\mathrm{AN}}^{(r)}$ denotes the restoration-stage AN power coefficient at the $l$-th AP, while $\mathcal L_{\mathrm{abs}}\subseteq\mathcal L$ denotes the local AP subset activated for rapid absorption. The absorption stage updates only $\mathcal L_{\mathrm{abs}}$ for rapid low-overhead mitigation, while the CPU simultaneously prepares the global restoration configuration shown in Fig.~\ref{fig1}; restoration then enables network-wide data/AN power control and PPZF topology update.

Let $S_{\mathrm{drop}} \triangleq 
\SSE_1^{(a)}\Big(\rho_{l,k}^{(a)} = \rho_{l,k}^{(n)\star}\Big)$ denote the SSE of $\UEo$ immediately after attack detection, assuming that the normal-stage power allocation is still in use (its definition is provided in subsection~\ref{sec:SSE}). For a prescribed performance absorption factor $\alpha\in(0,1)$, the absorption target is defined as
%--------------------------
\vspace{-0.2em}
\begin{equation}
S_{\alpha}^{\mathrm{abs}}
=
S_{\mathrm{drop}}+\alpha\big(S_{\mathrm N}^{\star}-S_{\mathrm{drop}}\big).
\label{eq:S_alpha_abs}
\end{equation}
The absorption stage stops once $\SSE_1^{(a)}$ first reaches $S_{\alpha}^{\mathrm{abs}}$, and the obtained point is then used to initialize the restoration stage. Both post-attack stages operate under the same contaminated CSI; the difference lies only in the admissible control actions.

%--------------------------
\vspace{-0.5em}
\subsection{Robust Worst-Case Model}
Under pilot spoofing, the received pilot signals associated with $\UEo$ are a combination of the legitimate and eavesdropping channels, making Eve's instantaneous CSI difficult to obtain at the APs. Therefore, following robust secure designs with bounded Eve-CSI uncertainty~\cite{EveCSI_Worst}, we adopt a worst-case model based on large-scale fading uncertainty. Specifically, let $\hat{\beta}_{l,e}$ denote the estimated large-scale fading coefficient associated with Eve at the $l$-th AP, and let $\epsilon_e$ denote the uncertainty coefficient. Hence, Eve's large-scale fading coefficient is assumed to lie in the uncertainty set $\beta_{l,e}\in\big[(1-\epsilon_{e})\hat{\beta}_{l,e},(1+\epsilon_{e})\hat{\beta}_{l,e}\big], ~ l\in\mathcal{L}$, and the corresponding upper bound is $\bar{\beta}_{l,e}=(1+\epsilon_{e})\hat{\beta}_{l,e}$.
% \begin{equation}
% \bar{\beta}_{l,e}=(1+\epsilon_{e})\hat{\beta}_{l,e}.
% \end{equation}

Using this bound, the corresponding variance term becomes
% \vspace{-0.1em}
\begin{equation}
\bar{\gamma}_{l,e}
=
\frac{\tau_p \rho_e \bar{\beta}_{l,e}^{2}}
{\tau_p \rho_p \beta_{l,1}+\tau_p \rho_e \bar{\beta}_{l,e}+1}.
\label{eq:gammae_bar_new}
\end{equation}
Here, the large-scale fading coefficient $\beta_{l,1}$ of the legitimate target user, i.e., $\UEo$, is assumed to be known at the APs through long-term channel statistics, while the uncertainty is imposed only on Eve-related large-scale fading. This worst-case model is considered in both the absorption and restoration stages.

%--------------------------
\vspace{-0.5em}
\subsection{PPZF Precoding and Protection Reconfiguration}\label{sec:ppzf}
%%%%%%%%%%%%%%%%%%%%%%%%%%%%%%%%%%%%%%%%%
%\vspace{-1em}
We adopt PPZF precoding at each AP, where each AP applies a partial zero-forcing (PZF) beamformer to a selected strong-user set and protective maximum-ratio transmission (PMRT) to the remaining weak users. Compared with classical ZF, PPZF avoids spending too many spatial degrees of freedom on interference suppression, while unlike MRT, it can still mitigate strong-user interference. Hence, PPZF provides a reasonable tradeoff between desired signal enhancement and interference control in CF-mMIMO systems~\cite{PPZF}.

For AP $l$, let $\mathcal{S}_l^{(s)} \subseteq \mathcal{K}$ denote the strong-user set in stage $s\in\{n,a,r\}$, and denote the weak-user set by $\mathcal{W}_l^{(s)} = \mathcal{K}\setminus \mathcal{S}_l^{(s)}$. For the normal benchmark topology, we define the baseline strong-user set $\widetilde{\mathcal{S}}_l$ according to the adopted PPZF strong-user construction rule based on the large-scale channel-gain information observed at AP $l$. In particular, the set is generated from the corresponding channel-gain metric while ensuring $|\widetilde{\mathcal{S}}_l|\le M-1$. This prevents the PZF from utilizing all $M$ antenna dimensions and preserves at least one residual spatial degree of freedom for PMRT or AN transmission. The remaining users are assigned to the baseline weak-user set $\widetilde{\mathcal{W}}_l=\mathcal{K}\setminus\widetilde{\mathcal{S}}_l$.

Define $\tau_{S,l}^{(s)} = |\mathcal{S}_l^{(s)}|$, and note that PPZF requires $\tau_{S,l}^{(s)} \le M-1$ in order to leave at least one residual degree of freedom for protection or AN. Based on the MMSE channel estimates in \eqref{eq:mmse_estimator}, define the estimated channel matrix at the $l$-th AP as $\hat{\qH}_l=[\hat{\qh}_{l,1},\ldots,\hat{\qh}_{l,K}] \in \mathbb{C}^{M\times K}$, and define the selection matrix $\qE_{\mathcal{S}_l^{(s)}} = [\qe_{i}]_{i\in \mathcal{S}_l^{(s)}} \in \mathbb{C}^{K\times \tau_{S,l}^{(s)}}$, where $\qe_i$ is the $i$-th column of $\qI_K$. Moreover, for $k\in \mathcal{S}_l^{(s)}$, define $\boldsymbol{\pi}_{l,k}^{(s)}=\qE_{\mathcal{S}_l^{(s)}}^\mathrm{H}\qe_k\in\mathbb{C}^{\tau_{S,l}^{(s)}\times1}$. The PPZF precoder used by the $l$-th AP  for $\UEk$ in stage $s$ is
%------------
%\vspace{-0.1em}
\begin{equation}
\qw_{l,k}^{(s)}
=
\begin{cases}
\qw_{l,k}^{\mathrm{PZF},(s)}, & k\in \mathcal{S}_l^{(s)},\\
\qw_{l,k}^{\mathrm{PMRT},(s)}, & k\in \mathcal{W}_l^{(s)}.
\end{cases}
\label{eq:ppzf_piecewise_new}
\end{equation}

For $k\in \mathcal{S}_l^{(s)}$, the PZF precoder is
% %---------------------
% \begin{equation}
% \qw_{l,k}^{\mathrm{PZF},(s)}
% =
% \frac{
% \hat{\qH}_l \qE_{\mathcal{S}_l^{(s)}}
% \Big(
% \qE_{\mathcal{S}_l^{(s)}}^{H}
% \hat{\qH}_l^{H}
% \hat{\qH}_l
% \qE_{\mathcal{S}_l^{(s)}}
% \Big)^{-1}
% \boldsymbol{\pi}_{l,k}^{(s)}
% }{
% \sqrt{
% \mathbb{E}\!\left\{
% \left\|
% \hat{\qH}_l \qE_{\mathcal{S}_l^{(s)}}
% \Big(
% \qE_{\mathcal{S}_l^{(s)}}^{H}
% \hat{\qH}_l^{H}
% \hat{\qH}_l
% \qE_{\mathcal{S}_l^{(s)}}
% \Big)^{-1}
% \boldsymbol{\pi}_{l,k}^{(s)}
% \right\|^2
% \right\}
% }
% }.
% \label{eq:pzf_precoder_new}
% \end{equation}
% %--------------------
\vspace{-0.3em}
\begin{equation}
\qw_{l,k}^{\mathrm{PZF},(s)}
=
\alpha_{\mathrm{PZF}}\hat{\qH}_l \qE_{\mathcal{S}_l^{(s)}}
\Big(
\qE_{\mathcal{S}_l^{(s)}}^{H}
\hat{\qH}_l^{H}
\hat{\qH}_l
\qE_{\mathcal{S}_l^{(s)}}
\Big)^{-1}
\boldsymbol{\pi}_{l,k}^{(s)},
\label{eq:pzf_precoder_new}
\end{equation}
%--------------------
where $\alpha_{\mathrm{PZF}}= \sqrt{(M-\tau_{S,l}^{(s)})\gamma_{l,k}^{(s)}}$ is the normalization factor, such that $\Ex\Big\{\big\Vert\qw_{l,k}^{\mathrm{PZF},(s)}\big\Vert^2\Big\}=1$.
% Its normalization term admits the closed-form expression
% \begin{equation}
% \mathbb{E}\!\left\{
% \left\|
% \hat{\qH}_l \qE_{\mathcal{S}_l^{(s)}}
% \Big(
% \qE_{\mathcal{S}_l^{(s)}}^{H}
% \hat{\qH}_l^{H}
% \hat{\qH}_l
% \qE_{\mathcal{S}_l^{(s)}}
% \Big)^{-1}
% \boldsymbol{\pi}_{l,k}^{(s)}
% \right\|^2
% \right\}
% =
% \frac{1}{(M-\tau_{S,l}^{(s)})\gamma_{l,k}^{(s)}}.
% \label{eq:pzf_norm_new}
% \end{equation}
Hence, for $k,t\in \mathcal{S}_l^{(s)}$, we have
%-------------------
\vspace{-0.7em}
\begin{equation}
\hat{\qh}_{l,k}^\mathrm{H}\qw_{l,t}^{\mathrm{PZF},(s)}
=
\begin{cases}
0, & t\neq k,\\
\sqrt{(M-\tau_{S,l}^{(s)})\gamma_{l,k}^{(s)}}, & t=k.
\end{cases}
\label{eq:pzf_gain_new}
\end{equation}
%-------------------

For $k\in \mathcal{W}_l^{(s)}$, define the null-space projector $\qB_l^{(s)}
=
\qI_M
-
\hat{\qH}_l \qE_{\mathcal{S}_l^{(s)}}
\Big(
\qE_{\mathcal{S}_l^{(s)}}^\mathrm{H}
\hat{\qH}_l^\mathrm{H}
\hat{\qH}_l
\qE_{\mathcal{S}_l^{(s)}}
\Big)^{-1}
\qE_{\mathcal{S}_l^{(s)}}^\mathrm{H}
\hat{\qH}_l^\mathrm{H}$.
% \begin{equation}
% \qB_l^{(s)}
% =
% \qI_M
% -
% \hat{\qH}_l \qE_{\mathcal{S}_l^{(s)}}
% \Big(
% \qE_{\mathcal{S}_l^{(s)}}^\mathrm{H}
% \hat{\qH}_l^\mathrm{H}
% \hat{\qH}_l
% \qE_{\mathcal{S}_l^{(s)}}
% \Big)^{-1}
% \qE_{\mathcal{S}_l^{(s)}}^\mathrm{H}
% \hat{\qH}_l^\mathrm{H}.
% \label{eq:B_proj_new}
% \end{equation}
Accordingly, the PMRT precoder is designed as $\qw_{l,k}^{\mathrm{PMRT},(s)}
=
\alpha_{\mathrm{PMRT}}\qB_l^{(s)}\hat{\qh}_{l,k}$,
where $\alpha_{\mathrm{PMRT}}\!=\!\frac{1}{\sqrt{(M-\tau_{S,l}^{(s)})\gamma_{l,k}^{(s)}}}$ to make $\Ex\Big\{\big\Vert \qw_{l,k}^{\mathrm{PMRT},(s)} \big\Vert^2\Big\}=1$.

Specifically, define the protection priority metric
%---------------------
\vspace{-0.2em}
\begin{equation}
\mu_l=\frac{\beta_{l,1}}{\bar{\beta}_{l,e}}, \quad l\in\mathcal{L}.
\label{eq:mu_l}
\end{equation}
Let $N_a$ and $N_p$ denote the numbers of APs selected for local absorption and PPZF reconfiguration, respectively. The set $\mathcal{P}_1$ contains the top-$N_p$ APs with the largest $\mu_l$.
 Then, for each $l\in\mathcal{P}_1$, the attacked user is forced into the recovered strong-user set:
%--------------------------
\vspace{-0.6em}
\begin{equation}
\mathcal{S}_l^{(r)}
=
\begin{cases}
\widetilde{\mathcal{S}}_l \cup \{1\}, & l\in\mathcal{P}_1,\ \tau_{S,l}^{(a)}<M-1,\ 1\notin\widetilde{\mathcal{S}}_l,\\[1mm]
\big(\widetilde{\mathcal{S}}_l\setminus\{k_l^\star\}\big)\cup\{1\}, & l\in\mathcal{P}_1,\ \tau_{S,l}^{(a)}=M-1,\ 1\notin\widetilde{\mathcal{S}}_l,\\[1mm]
\widetilde{\mathcal{S}}_l, & \text{otherwise},
\end{cases}
\label{eq:Sr_new}
\end{equation}
where $k_l^\star = \arg\min_{k\in\widetilde{\mathcal{S}}_l}\beta_{l,k}$. 

In the absorption stage, the baseline topology is retained, i.e., $\mathcal S_l^{(n)}=\mathcal S_l^{(a)}=\widetilde{\mathcal S}_l$. In the restoration stage, a PPZF reconfiguration is activated by selecting the top-$N_p$ APs according to \eqref{eq:mu_l} and forcing $\UEo$ into their strong-user sets. 
% The main restoration gain comes from this topology update, while restoration-stage AN is treated only as an auxiliary enhancement under contaminated CSI.

%--------------------------
\vspace{-1em}
\subsection{Achievable SE and SSE with Stage-Dependent PPZF}\label{sec:SSE}
%----------------------------
For stage $s\in\{n,a,r\}$, let $\mathcal{S}_l^{(s)}$ denote the strong-user set at AP $l$, and define
%---------------------
\vspace{-0.2em}
\begin{equation}
\delta_{l,k}^{(s)}=
\begin{cases}
1, & k\in \mathcal{S}_l^{(s)},\\
0, & k\notin \mathcal{S}_l^{(s)},
\end{cases}
~
\tau_{S,l}^{(s)} = |\mathcal{S}_l^{(s)}|,
~
N_{l,\mathrm{AN}}^{(s)} = M-\tau_{S,l}^{(s)}.
\label{eq:delta_tau_nan}
\end{equation}
Since $\tau_{S,l}^{(r)}\le M-1$, there remains at least one residual spatial dimension for activated AN. The transmitted signal at AP $l$ in stage $s$ is written as
%--------------------------
\vspace{-0.3em}
\begin{equation}
\qx_l^{(s)}
=
\sum\nolimits_{k\in\mathcal{K}}\sqrt{\rho_{l,k}^{(s)}}\,\qw_{l,k}^{(s)} q_k
+
\qP_{l,\mathrm{AN}}^{(s)}\qz_l^{(s)},
\label{eq:tx_signal_no_switch}
\end{equation}
where $q_k\sim\mathcal{CN}(0,1)$ is the information symbol for $\UEk$, 
$\qz_l^{(s)}\sim\mathcal{CN}\!\big(\mathbf{0},\rho_{l,\mathrm{AN}}^{(s)}\qI_{N_{l,\mathrm{AN}}^{(s)}}\big)$ is the AN vector, and $\qP_{l,\mathrm{AN}}^{(s)}\in\mathbb{C}^{M\times N_{l,\mathrm{AN}}^{(s)}}$ spans the residual null space left by the PPZF protected-user subspace. Hence, the AN term is inactive whenever $\rho_{l,\mathrm{AN}}^{(s)}=0$ or $N_{l,\mathrm{AN}}^{(s)}=0$.

% For the legitimate users, define the effective PPZF coefficients
% \begin{equation}
% a_{l,k}^{(s)}=\sqrt{\left(M-\tau_{S,l}^{(s)}\right)\gamma_{l,k}^{(s)}},
% \qquad
% b_{l,k}^{(s)}=\beta_{l,k}-\delta_{l,k}^{(s)}\gamma_{l,k}^{(s)}.
% \label{eq:ab_legitimate}
% \end{equation}
% In addition, using the worst-case Eve parameters $\bar{\beta}_{l,e}$ and $\bar{\gamma}_{l,e}$, define
% \begin{equation}
% c_{l}^{(s)}=\sqrt{\left(M-\tau_{S,l}^{(s)}\right)\bar{\gamma}_{l,e}},
% \qquad
% d_{l}^{(s)}=\bar{\beta}_{l,e}-\delta_{l,1}^{(s)}\bar{\gamma}_{l,e}.
% \label{eq:cd_eve}
% \end{equation}

For the normal and absorption stages, $\rho_{l,\mathrm{AN}}^{(n)}=\rho_{l,\mathrm{AN}}^{(a)}=0$. Let $a_{l,k}^{(s)}=\sqrt{\big(M-\tau_{S,l}^{(s)}\big)\gamma_{l,k}^{(s)}}$ and $b_{l,k}^{(s)}=\beta_{l,k}-\delta_{l,k}^{(s)}\gamma_{l,k}^{(s)}$. Then, the expression of the achievable signal-to-interference-plus-noise ratio (SINR) for $\UEk$ in stage $s$ is
\begin{equation}
\SINR_{k}^{(s)}
=
\frac{
\Big(
\sum\nolimits_{l\in\mathcal{L}} a_{l,k}^{(s)}\sqrt{\rho_{l,k}^{(s)}}
\Big)^2
}{
\sum\nolimits_{i\in\mathcal{K}}\sum\nolimits_{l\in\mathcal{L}} b_{l,k}^{(s)}\rho_{l,i}^{(s)}
+
\sum\nolimits_{l\in\mathcal{L}} N_{l,\mathrm{AN}}^{(s)} b_{l,k}^{(s)} \rho_{l,\mathrm{AN}}^{(s)}
+1
}.
\label{eq:sinr_user_stage}
\end{equation}
%---------------

Using $\bar{\beta}_{l,e}$ and $\bar{\gamma}_{l,e}$, the worst-case effective SINR of Eve in stage $s$ is expressed as
%----------------------------------
\begin{equation}
\SINR_{e}^{(s)}
=
\frac{
\Big(
\sum\nolimits_{l\in\mathcal{L}} c_{l}^{(s)}\sqrt{\rho_{l,1}^{(s)}}
\Big)^2
+
\sum\nolimits_{l\in\mathcal{L}} d_{l}^{(s)}\rho_{l,1}^{(s)}
}{
\sum_{i=2}^{K}\sum\nolimits_{l\in\mathcal{L}} d_{l}^{(s)}\rho_{l,i}^{(s)}
+
\sum\nolimits_{l\in\mathcal{L}} N_{l,\mathrm{AN}}^{(s)} d_{l}^{(s)} \rho_{l,\mathrm{AN}}^{(s)}
+1
},
\label{eq:sinr_eve_stage}
\end{equation}
%-----------------
where $c_{l}^{(s)}=\sqrt{\left(M-\tau_{S,l}^{(s)}\right)\bar{\gamma}_{l,e}}$ and $d_{l}^{(s)}=\bar{\beta}_{l,e}-\delta_{l,1}^{(s)}\bar{\gamma}_{l,e}$.

Accordingly, the SE of $\UEk$ and the worst-case SE of Eve are respectively given by
\begin{equation}
\SE_{k}^{(s)}\!\!=\!\log_2\!\Big(1\!+\!\SINR_{k}^{(s)}\Big),
\,\SE_{e}^{(s)}\!\!= \!\log_2\!\Big(1\!+\!\SINR_{e}^{(s)}\Big).
\label{eq:se_defs_stage}
\end{equation}
Finally, the attacked-user SSE is
\begin{equation}
\SSE_{1}^{(s)}
\!=\!
\left[
\SE_{1}^{(s)}-\SE_{e}^{(s)}
\right]^+
\!=\!
\left[
\log_2\!\left(
\frac{1+\SINR_{1}^{(s)}}{1+\SINR_{e}^{(s)}}
\right)
\right]^+.
\label{eq:sse_stage}
\end{equation}

%============================
%\vspace{-0.8em}
\section{Problem Formulation and Solution}
%============================
To explicitly distinguish the normal benchmark generation from the post-disruption resilient control, we formulate the design over two time scales.

%============================
\vspace{-0.5em}
\subsection{Normal-Stage Problem}
%============================
During the normal phase, there is no eavesdropping attack. We consider a max-min fairness (MMF) problem that aims to maximize the users’ SE while ensuring all users receive a guaranteed minimum quality-of-service (QoS). Let $\rho_{l,k}^{(n)}$ denote the normal-phase data power coefficient. The normal stage problem is formulated as
\begin{subequations}
\begin{align}
\max_{\rho_{l,k}^{(n)}}
\quad
& \min_{k \in \mathcal{K}} \, \SE_k^{(n)}
\label{prob:PN_obj}\\
\text{s.t.}\quad
& \SE_k^{(n)} \ge \SE_{k,\min},~\forall k\in\mathcal{K},
\label{prob:PN_qos}\\
& \sum\nolimits_{k\in\mathcal{K}}\rho_{l,k}^{(n)} \le P_l^{\max},~ \forall l\in\mathcal{L},
\label{prob:PN_power}\\
& \rho_{l,k}^{(n)} \ge 0, ~ \forall l\in\mathcal{L},~ k\in\mathcal{K},
\label{prob:PN_nonneg}
\end{align}\label{PN}
\end{subequations}
\!\!where $\SE_{k,\min}$ denotes the minimum SE requirement of $\UEk$, and $P_l^{\max}$ denotes the maximum transmit power at AP $l$. Let $\rho_{l,k}^{(n)\star}$ denote the local optimal power allocation obtained from \eqref{PN}. After solving the normal-stage fair allocation problem, the service level of $\UEo$ is extracted as
\begin{equation}
    S_{\mathrm N}^{\star} \triangleq\SE_1^{(n)}\Big(\rho_{l,k}^{(n)\star}\Big)=\SSE_1^{(n)}\Big(\rho_{l,k}^{(n)\star}\Big),
\end{equation}
which is used as the pre-attack secrecy benchmark since no active eavesdropping is present.

%============================
\vspace{-0.5em}
\subsection{Absorption-Stage Problem}
%============================
Following the normal phase, upon receiving a pilot-spoofing alarm, the system switches to the absorption phase. Unlike a conventional attacked-state redesign over the network, the absorption stage is configured as a limited and rapid operation. Specifically, we use the local absorption priority metric in \eqref{eq:mu_l} to construct the local absorption AP subset $\mathcal L_{\mathrm{abs}}\subseteq\mathcal L$ by selecting the top-$N_a$ APs with the largest $\mu_l$. Hence, the absorption controller only updates the APs that are most favorable to the attacked user relative to the worst-case Eve. In this stage, the normal-stage PPZF topology is retained. For APs outside this subset, the power allocation remains unchanged, i.e., 
%the normal phase, the system detects a pilot-spoofing attack and switches to the absorption phase. Unlike a conventional attacked-state redesign over the whole network, the absorption stage is configured as a limited and rapid operation. Specifically, we use the local absorption priority metric in \eqref{eq:mu_l} to construct the local absorption AP subset $\mathcal L_{\mathrm{abs}}\subseteq\mathcal L$ by selecting the top-$N_a$ APs with the largest $\mu_l$. Hence, the absorption controller only updates the APs that are most favorable to the attacked user relative to the worst-case Eve. In this stage, the normal-stage PPZF topology is retained. For APs outside this subset, the power allocation remains unchanged, i.e.,
\begin{equation}
\rho_{l,k}^{(a)}=\rho_{l,k}^{(n)\star},  ~ l\notin \mathcal L_{\mathrm{abs}},~\forall k\in\mathcal K,
\label{eq:abs_fixed_outside}
\end{equation}
where $\rho_{l,k}^{(n)\star}$ is the optimized normal-stage power allocation. Hence, the absorption-stage design is formulated as
%--------------------
\vspace{-0.1em}
\begin{subequations}
\begin{align}
\max_{\rho_{l,k}^{(a)}}
\quad
& \SSE_{1}^{(a)}
\\
\text{s.t.}\quad
& \SE_{k}^{(a)} \ge \SE_{k,\min}, ~ k=2,\ldots,K,
\\
& \sum\nolimits_{k\in\mathcal{K}}\rho_{l,k}^{(a)} \le P_{l}^{\max}, ~ l\in\mathcal{L},
\\
& \rho_{l,k}^{(a)}\ge 0,~ \forall l\in\mathcal{L},\ k\in\mathcal{K},
\\
& \rho_{l,k}^{(a)}=\rho_{l,k}^{(n)\star}, ~l\notin \mathcal L_{\mathrm{abs}},~\forall k\in\mathcal K.
\end{align}
\label{PA}
\end{subequations}
\!\!Problem \eqref{PA} is solved only over the variable subset 
$\rho_{l,k}^{(a)},~l\in\mathcal L_{\mathrm{abs}},~k\in\mathcal K$, while the powers of APs outside $\mathcal L_{\mathrm{abs}}$ are fixed at their normal-stage values. The absorption stage terminates when $\SSE_1^{(a)}\ge S_{\alpha}^{\mathrm{abs}}$ or the maximum number of iterations is reached.
%============================
\vspace{-1.7em}
\subsection{Restoration-Stage Problem}
%============================
Let $\boldsymbol{\rho}_{\alpha}^{(a)}=\big\{\rho_{\alpha,l,k}^{(a)}\big\}$ denote the absorbed attacked-stage power allocation returned by the absorption stage. Starting from this absorbed power allocation, the restoration stage optimizes all APs under the richer action space $\mathcal A_{\mathrm{rec}}$, defined in~\eqref{A_rec}, 
%----------
\vspace{-0.1em}
\begin{subequations}
\begin{align}
\max_{\rho_{l,k}^{(r)},\,\rho_{l,\mathrm{AN}}^{(r)}}
\quad
& \SSE_{1}^{(r)}\\
\text{s.t.}\quad
& \SE_{k}^{(r)} \ge \SE_{k,\min}, ~ k=2,\ldots,K,\\
& \sum\nolimits_{k\in\mathcal{K}}\rho_{l,k}^{(r)}
+
N_{l,\mathrm{AN}}^{(r)}\rho_{l,\mathrm{AN}}^{(r)}
\le P_l^{\max}, ~ l\in\mathcal{L},\\
& \rho_{l,k}^{(r)}\ge 0,~
\rho_{l,\mathrm{AN}}^{(r)}\ge 0.
\end{align}\label{PR}
\end{subequations}

%============================
\vspace{-1.9em}
\subsection{SCA Reformulation for Multi-Stage Optimization}
%============================
The normal-stage benchmark problem \eqref{PN} and the post-attack problems \eqref{PA}--\eqref{PR} are non-convex due to the coupled SINR terms. However, after fixing the stage-dependent topology and the corresponding coefficient sets, they admit a unified SCA treatment.
For each stage $(s)$, let $\qu_{k}^{(s)}=\Big[\sqrt{\rho_{1,k}^{(s)}},\ldots,\sqrt{\rho_{L,k}^{(s)}}\Big]^{\mathrm T}$ and  define $\qa_{k}^{(s)}=
\big[a_{1,k}^{(s)},\ldots,a_{L,k}^{(s)}\big]^{\mathrm T},
\quad
\qA_{k}^{(s)}=\mathrm{diag}\big(b_{1,k}^{(s)},\ldots,b_{L,k}^{(s)}\big)$, $\qc^{(s)}=
[c_{1}^{(s)},\ldots,c_{L}^{(s)}]^{\mathrm T},
\quad
\qD^{(s)}=\mathrm{diag}\big(d_{1}^{(s)},\ldots,d_{L}^{(s)}\big)$. Then, the legitimate-user SINR, in~\eqref{eq:sinr_user_stage}, can be written as
%-----------------------------
\begin{equation}
\SINR_{k}^{(s)}
=
\frac{\big(\qa_{k}^{(s)\mathrm T}\qu_{k}^{(s)}\big)^2}{y_k^{(s)}},
\label{eq:sinr_user_compact_final}
\end{equation}
where $y_k^{(s)}
\!=\!
1\!+\!
\sum\nolimits_{i\in\mathcal{K}}\big \|\qA_{k}^{(s)1/2}\qu_{i}^{(s)}\big\|_2^2
\!+\!
\sum\nolimits_{l\in\mathcal{L}}N_{l,\mathrm{AN}}^{(s)}\,b_{l,k}^{(s)}\,\rho_{l,\mathrm{AN}}^{(s)}$.
% %------------
% \vspace{-0.2em}
% \begin{align*}
% y_k^{(s)}
% =
% 1+
% \sum\nolimits_{i\in\mathcal{K}}\big \|\qA_{k}^{(s)1/2}\qu_{i}^{(s)}\big\|_2^2
% +
% \sum\nolimits_{l\in\mathcal{L}}N_{l,\mathrm{AN}}^{(s)}\,b_{l,k}^{(s)}\,\rho_{l,\mathrm{AN}}^{(s)}.  
% \end{align*}
% %-----------------

Similarly, the worst-case SINR of Eve is expressed as
\begin{equation}
\SINR_{e}^{(s)}
=
\frac{
\big(\qc^{(s)\mathrm T}\qu_{1}^{(s)}\big)^2
+
\big\|\qD^{(s)1/2}\qu_{1}^{(s)}\big\|_2^2
}{
z^{(s)}
},
\label{eq:sinr_eve_compact_final}
\end{equation}
with $z^{(s)}
=
1+
\sum_{i=2}^{K}\big\|\qD^{(s)1/2}\qu_{i}^{(s)}\big\|_2^2
+
\sum\nolimits_{l\in\mathcal{L}}N_{l,\mathrm{AN}}^{(s)}\,d_l^{(s)}\,\rho_{l,\mathrm{AN}}^{(s)}$.

At SCA iteration $\nu$, introduce variables $x_k^{(s)}$ satisfying $\qa_{k}^{(s)\mathrm T}\qu_{k}^{(s)} \ge x_k^{(s)},
~ k\in\mathcal{K}$, and define the denominator functions
\begin{align}
\phi_k^{(s)}\!\big(\qu_i^{(s)},\rho_{\mathrm{AN}}^{(s)}\big)
=&
1+
\sum\nolimits_{i\in\mathcal{K}}\big\|\qA_{k}^{(s)1/2}\qu_{i}^{(s)}\big\|_2^2
\nonumber\\
&
+
\sum\nolimits_{l\in\mathcal{L}}N_{l,\mathrm{AN}}^{(s)}\,b_{l,k}^{(s)}\,\rho_{l,\mathrm{AN}}^{(s)}.
\label{eq:phi_k_final}
\end{align}

Using the first-order global lower bound of the convex quadratic-over-linear function,
%----------------------------
\vspace{-0.1em}
\begin{equation}
\frac{x^2}{y}
\ge
\frac{2\bar{x}}{\bar{y}}x
-
\left(\frac{\bar{x}}{\bar{y}}\right)^2 y,
~ x>0,\ y>0,
\label{eq:frac_lb_final}
\end{equation}
the SINR of $\UEk$ admits the affine lower bound
\begin{equation}
\SINR_{k}^{\mathrm{lb},(s,\nu)}
\!=\!
\frac{2\bar{x}_{k}^{(s,\nu)}}{\bar{y}_{k}^{(s,\nu)}}x_k^{(s)}
\!-\!
\left(
\frac{\bar{x}_{k}^{(s,\nu)}}{\bar{y}_{k}^{(s,\nu)}}
\right)^2\!\!
\phi_k^{(s)}\!\Big(\{\qu_i^{(s)}\},\rho_{\mathrm{AN}}^{(s)}\Big),
\label{eq:uk_lb_final}
\end{equation}
where $\bar{x}_{k}^{(s,\nu)}$ and $\bar{y}_{k}^{(s,\nu)}$ are evaluated at the previous iteration.

For Eve's term, introduce an auxiliary variable $v_e^{(s)}$ satisfying $\SINR_{e}^{(s)} \le v_e^{(s)}$. Then, define the affine lower bound of Eve's denominator as
\begin{align}
\underline{z}^{(s,\nu)}
&\!=\!
1\!+\!
\sum\nolimits_{i=2}^{K}
\left[
2\Big(\qu_{i}^{(s,\nu)}\Big)^{\mathrm T}\qD^{(s)}\qu_{i}^{(s)}
\!-\!
\Big(\qu_{i}^{(s,\nu)}\Big)^{\mathrm T}\qD^{(s)}\qu_{i}^{\!(s,\nu)}
\!\right]
\nonumber\\
&
\!+
\sum\nolimits_{l\in\mathcal{L}}N_{l,\mathrm{AN}}^{(s)}\,d_l^{(s)}\,\rho_{l,\mathrm{AN}}^{(s)}.
\label{eq:ze_lb_final}
\end{align}
To this end, the upper-bounding condition for Eve can be equivalently imposed through the rotated second-order cone (SOC) constraint
\begin{equation}
\left\|
\begin{bmatrix}
2\qc^{(s)\mathrm T}\qu_{1}^{(s)}\\
2\qD^{(s)1/2}\qu_{1}^{(s)}\\
v_e^{(s)}-\underline{z}^{(s,\nu)}
\end{bmatrix}
\right\|_2
\le
v_e^{(s)}+\underline{z}^{(s,\nu)}.
\label{eq:eve_soc_final}
\end{equation}

Moreover, since $\log_2\big(1+v_e^{(s)}\big)$ is concave, its first-order Taylor expansion around the previous iteration $\bar v_e^{(s,\nu)}$ yields the global upper bound
%----------------------------
\vspace{-0.5em}
\begin{equation}
\widehat g_e^{(s,\nu)}(v_e^{(s)})
=
\log_2\big(1+\bar v_e^{(s,\nu)}\big)
+
\frac{v_e^{(s)}-\bar v_e^{(s,\nu)}}{(1+\bar v_e^{(s,\nu)})\ln 2}.
\label{eq:eve_log_ub_final}
\end{equation}

For all stages, the coupled SINR terms are solved via a unified SCA framework. At iteration $\nu$, the legitimate-user SINR is lower-bounded by applying \eqref{eq:frac_lb_final}, which yields a concave lower-bound surrogate $\SINR_{k}^{\mathrm{lb},(s,\nu)}$ for each user. For the post-attack secrecy stages, an auxiliary variable $v_e^{(s)}$ is introduced for Eve's term, and the corresponding upper-bounding condition is enforced through the SOC constraint in \eqref{eq:eve_soc_final}. Hence, the normal benchmark problem and the two post-attack problems are solved iteratively via convex SCA subproblems under their stage-dependent feasible sets. The absorption and restoration stages share the same SCA machinery, while differing in topology, AN availability, and optimized AP set. The overall two-timescale robust security-resilient design is summarized in \textbf{Algorithm~\ref{alg1}}.

\textbf{\textit{Complexity Analysis: }}The long-term normal benchmark has complexity \(\mathcal O(I_{\mathrm{norm}}(LK)^3)\), where $I_{\mathrm{norm}}$ denotes the number of SCA iterations in the normal stage. While the absorption stage only optimizes the local AP subset $\mathcal L_{\mathrm{abs}}$ with $|\mathcal L_{\mathrm{abs}}|=N_a \ll L$, its per-iteration complexity scales with the reduced variable dimension associated with $N_aK$. By contrast, the restoration stage operates over the full AP set. Therefore, the total short-term complexity is $\mathcal{O}\!\left(
I_{\mathrm{abs}}(N_aK)^3
+
I_{\mathrm{res}}((K+1)L)^3
\right)$, 
where $I_{\mathrm{abs}}$ and $I_{\mathrm{res}}$ denote the number of SCA iterations in the absorption and restoration stages, respectively. Accordingly, absorption immediately updates only $N_aK$ data-power coefficients, reducing short-term control signaling relative to full-network restoration.

\begin{algorithm}[t]
\caption{Two-Timescale Robust Security-Resilient Design}
\label{alg1}
\begin{algorithmic}[1]
\STATE \textbf{Input:} $\beta_{l,k}$,  $\hat{\beta}_{l,e}$, $\epsilon_e$, $\rho_p$, $\rho_e$, $\alpha$, $N_a$, $N_p$, $T_{\max}^{(a)}$, $T_{\max}^{(r)}$.
\STATE \textbf{Normal stage:}
\STATE Solve \eqref{PN} to obtain $\rho_{l,k}^{(n)\star}$ and $S_{\mathrm N}^{\star}$.
\STATE \textbf{Absorption stage:}
\STATE Compute $\mu_l$ from \eqref{eq:mu_l} and construct $\mathcal L_{\mathrm{abs}}$ from the top-$N_a$ APs.
\STATE Set $\mathcal S_l^{(a)}=\widetilde{\mathcal S}_l$, $\rho_{l,\mathrm{AN}}^{(a)}=0$, and $\rho_{l,k}^{(a,0)}=\rho_{l,k}^{(n)\star}$.
\STATE Compute $S_{\mathrm{drop}}$, $S_{\alpha}^{\mathrm{abs}}$ from \eqref{eq:S_alpha_abs} and set $t^{(a)}=0$.
\REPEAT
    \STATE Solve \eqref{PA} over $l\in\mathcal L_{\mathrm{abs}}$ to update $\rho_{l,k}^{(a)}$ and evaluate $\SSE_1^{(a)}$.
    \STATE $t^{(a)}= t^{(a)}+1$.
\UNTIL{$\SSE_1^{(a)}\ge S_{\alpha}^{\mathrm{abs}}$ or $t^{(a)}\ge T_{\max}^{(a)}$}
% \STATE \textbf{Output:} $\rho_{\alpha,l,k}^{(a)}$.
\STATE \textbf{Restoration stage:}
\STATE Construct $\mathcal P_1$ from the top-$N_p$ APs according to $\mu_l$.
\STATE Construct the recovered PPZF topology $\mathcal S_l^{(r)}$ from \eqref{eq:Sr_new}.
\STATE Initialize $\rho_{l,k}^{(r,0)}=\rho_{l,k}^{(a)\star}$, $\rho_{l,\mathrm{AN}}^{(r,0)}\ge 0$ and $t^{(r)}=0$.
\REPEAT
\STATE Solve \eqref{PR} to obtain $\rho_{l,k}^{(r)\star}$ and $\rho_{l,\mathrm{AN}}^{(r)\star}$.
\STATE $t^{(r)}=t^{(r)}+1.$
\UNTIL{\(t^{(r)}\ge{T_{\max}^{(r)}}\)}
\STATE \textbf{Output:} $S_{\mathrm N}^{\star}$, $\rho_{l,k}^{(n)\star}$, $\rho_{l,k}^{(a)\star}$, $\rho_{l,k}^{(r)\star}$, $\rho_{l,\mathrm{AN}}^{(r)\star}$.
\end{algorithmic}
\end{algorithm}
%-------
\setlength{\textfloatsep}{0.1cm}
%----------------------------

%---------------------------------
\vspace{-0.8em}
\section{Numerical Results}
%----------------------------
\vspace{0.2em}
\begin{figure*}[t]
%\vspace{-0.2cm}
\centering
\begin{minipage}[t]{0.32\textwidth}
\centering
\captionsetup{type=figure,font=small,format=plain,justification=justified,singlelinecheck=false,width=\linewidth}
\includegraphics[width=\linewidth]{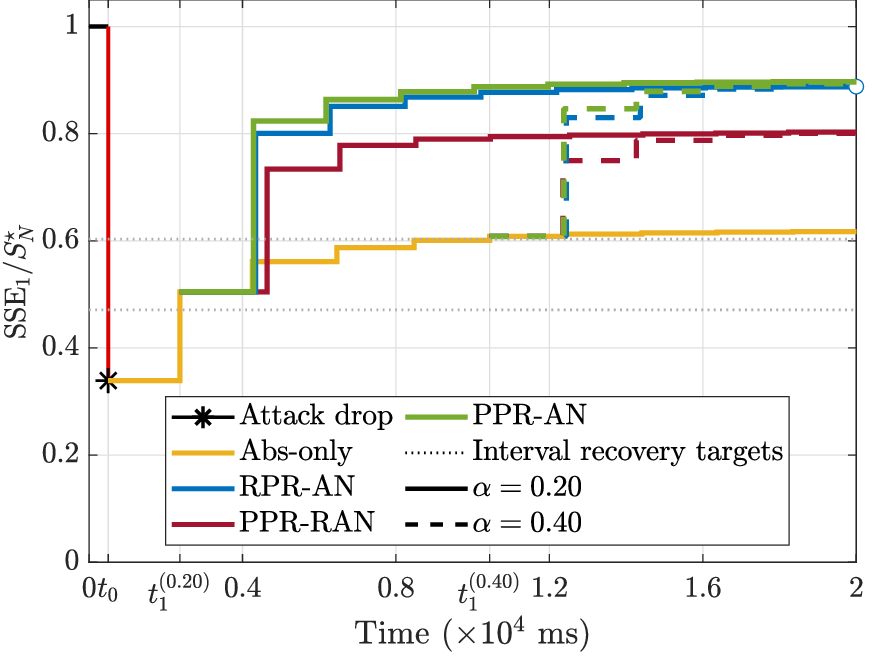}
\vspace{-1.7em}
\captionof{figure}{Post-attack recovery trajectories of $\mathrm{SSE}_1$ under different schemes.}
\label{Fig.conv}
\end{minipage}
\hfill
\begin{minipage}[t]{0.32\textwidth}
\centering
\captionsetup{type=figure,font=small,format=plain,justification=justified,singlelinecheck=false,width=\linewidth}
\includegraphics[width=\linewidth]{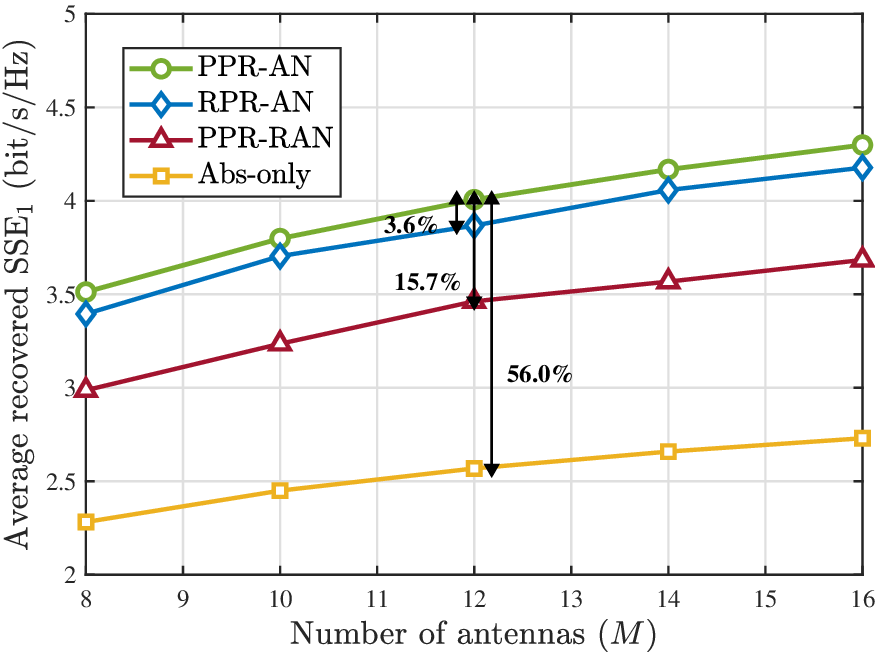}
\vspace{-1.7em}
\captionof{figure}{Impact of the AP antenna number on the recovered $\mathrm{SSE}_1$ ($L=20$).}
\label{Fig.M}
\end{minipage}
\hfill
\begin{minipage}[t]{0.32\textwidth}
\centering
\captionsetup{type=figure,font=small,format=plain,justification=justified,singlelinecheck=false,width=\linewidth}
\includegraphics[width=\linewidth]{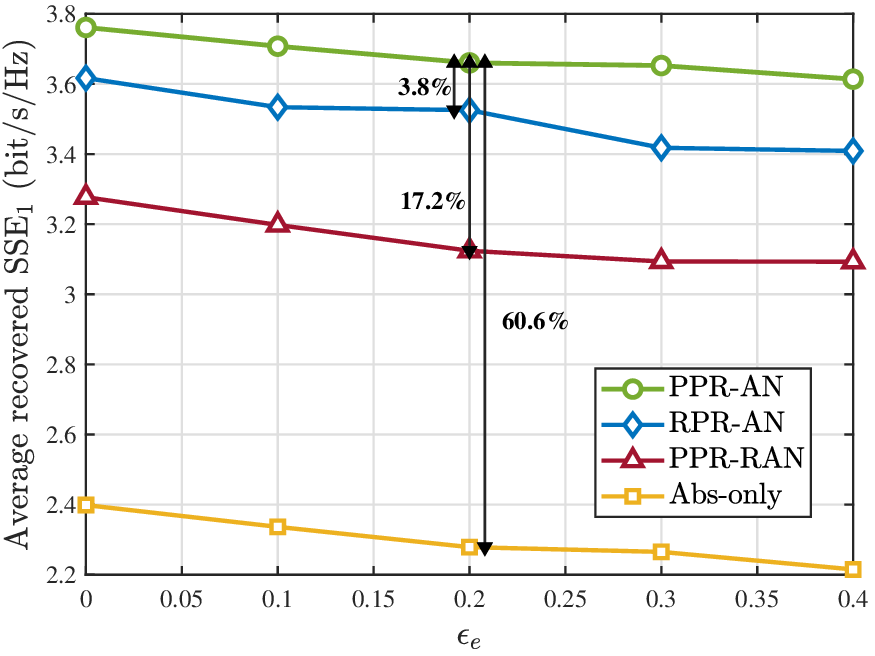}
\vspace{-1.7em}
\captionof{figure}{Impact of uncertainty coefficient $\epsilon_e$ on the recovered $\mathrm{SSE}_1$ ($M=8$, $L=20$).}
\label{Fig.epsilon}
\end{minipage}
\vspace{-1.7em}
\end{figure*}
%%%%%%%%%%%%%%%%%%%%%%%%%%%%%%%%%

We evaluate the proposed robust security-resilient design in CF-mMIMO systems under active pilot spoofing. The large-scale fading coefficients are generated using the simulation setup in~\cite{Atiya:TWC:2024}. Unless otherwise specified, we set $L=20$, $M=8$, $K=10$, $\alpha=0.3$, $N_a=3$, and $N_p=4$, where Eve attacks the target user $\UEo$ by transmitting the same pilot. The proposed PPZF reconfiguration (PPR) scheme with optimized AN, denoted \textbf{PPR-AN}, is compared with three benchmarks: \textbf{RPR-AN}, which randomly selects the protected APs while optimizing AN power; \textbf{PPR-RAN}, which uses the proposed protection set with random AN directions; and \textbf{Abs-only}, which stops after absorption-stage power optimization without restoration.

Fig.~\ref{Fig.conv} shows the recovery trajectories of the attacked user’s normalized secrecy performance under different schemes. After the attack alarm at $t_0=500$ ms, $\mathrm{SSE}_1$ drops from its normal stage to a loss level. The system then enters the absorption stage, where local power reallocation is applied over an $N_a=3$ AP subset.\footnote{Based on simulations, the chosen $N_a$ achieves a desirable balance between computational complexity and performance, while its optimal selection remains an interesting direction for future work.} Two absorption targets, $\alpha=0.2$ and $\alpha=0.4$, represent prescribed loss levels and initialize the restoration stage. The \textbf{Abs-only} scheme provides limited recovery, as it relies on local power reallocation without topology reconfiguration. In contrast, all restoration schemes improve secrecy performance once the absorption stage reaches the $\alpha$ target. Among them, the \textbf{PPR-AN} scheme achieves the highest restored level, demonstrating the benefit of combining PPZF reconfiguration with AN-assisted recovery. The two absorption targets highlight the time–quality tradeoff: a smaller $\alpha$ triggers restoration earlier, whereas a larger $\alpha$ allows more local recovery during restoration preparation.

Fig.~\ref{Fig.M} shows the average recovered $\mathrm{SSE}_1$ versus the number of antennas $M$. Increasing $M$ provides all schemes with more spatial degrees of freedom, enhancing the useful signal and suppressing interference in both stages. However, gains are clearly scheme-dependent. The absorption-only scheme remains the weakest across the entire range of $M$, since it is limited to short-horizon local power reallocation and cannot exploit network-wide post-attack restoration. The two restoration benchmarks achieve higher recovered secrecy, confirming the need to switch from local reallocation to global restoration in the proposed two-timescale design. Among all schemes, \textbf{PPR-AN} consistently achieves the highest recovered $\mathrm{SSE}_1$. At $M=12$, it improves the recovered $\mathrm{SSE}_1$ by about $3.6\%$, $15.7\%$, and $56\%$ over \textbf{RPR-AN}, \textbf{PPR-RAN}, and \textbf{Abs-only}, respectively. Hence, the \textbf{PPR-AN} utilizes the increased spatial resources more effectively than random protection or random AN strategies.

% Fig.~\ref{Fig.M} shows the average recovered $\mathrm{SSE}_1$ versus the number of antennas $M$. As $M$ increases, all schemes benefit from the enlarged spatial degrees of freedom, which improve useful-signal enhancement and interference suppression in both stages. However, the performance gain is clearly scheme-dependent. The absorption-only scheme remains the weakest over the whole range of $M$, since it is restricted to short-horizon local power reallocation and cannot exploit network-wide restoration after the attack. The two restoration benchmarks achieve better recovered secrecy, which confirms the necessity of switching from local reallocation to global restoration in the proposed two-timescale design. Among all schemes, \textbf{PPR-AN} consistently achieves the highest recovered $\mathrm{SSE}_1$. For example, at $M=12$, it improves the recovered $\mathrm{SSE}_1$ by about $3.6\%$, $15.7\%$, and $56.0\%$ over \textbf{RPR-AN}, \textbf{PPR-RAN}, and \textbf{Abs-only}, respectively. This shows that the \textbf{proposed PPZF reconfiguration (PPR)} together with optimized AN can utilize the increased spatial resources more effectively than random protection or random AN strategies.

Fig.~\ref{Fig.epsilon} shows the average recovered $\mathrm{SSE}_1$ versus the uncertainty coefficient $\epsilon_e$, which determines the worst-case bound adopted for Eve-related large-scale fading in the robust design. As $\epsilon_e$ increases, the recovered secrecy performance of all schemes gradually decreases, since a larger $\epsilon_e$ corresponds to a more conservative worst-case model, which enlarges the presumed threat region of Eve and therefore tightens the secure transmission design in both the absorption and restoration stages. Moreover, over the whole range of $\epsilon_e$, the \textbf{PPR-AN} scheme consistently outperforms the two random benchmarks (\textbf{RPR-AN} and \textbf{PPR-RAN}). For example, at $\epsilon_e=0.2$, it achieves about $3.8\%$, $17.2\%$, and $60.6\%$ higher recovered $\mathrm{SSE}_1$ than \textbf{RPR-AN}, \textbf{PPR-RAN}, and \textbf{Abs-only}, respectively. Overall, even as the uncertainty of large-scale fading increases, the proposed robust multi-stage framework maintains its secrecy recovery advantage.
% Fig.~\ref{Fig.r} shows the average recovered $\mathrm{SSE}_1$ versus the distance $r_{e,1}$ between the Eve and the targeted user. As expected, the recovered secrecy performance improves for all schemes as $r_{e,1}$ increases, because a larger separation weakens both the pilot spoofing impact and the effective downlink leakage toward the Eve. More importantly, the ordering of the curves remains unchanged over the whole range, with \textbf{proposed PPR + AN} always achieving the best recovery result. This indicates that the gain of the proposed method is not tied to a specific threat location, but remains effective under different spatial attack conditions. Taken together, Figs.~\ref{Fig.M} and \ref{Fig.r} show that the proposed framework maintains its advantage under both varying spatial processing capability and varying threat proximity, which supports the robustness of the proposed two-timescale resilience design.

\vspace{-0.7em}
\section{Conclusion}
This paper proposed a robust security-resilient transmission framework for CF-mMIMO systems under active pilot spoofing. A normal benchmark was first established under no attack. After attack detection, the system sequentially performs an absorption stage for fast secure fallback and a restoration stage for further secrecy recovery. To support this design, a worst-case large-scale uncertainty model for Eve and a protection-oriented PPZF reconfiguration with restoration-stage AN power optimization were developed. Numerical results showed that the proposed sequential design effectively improves both secure absorption and secure restoration under contaminated CSI.

\vspace{-0.3em}
\bibliographystyle{IEEEtran}
\bibliography{IEEEabrv,Reference}

\end{document}